# Large magnetoresistance from long-range interface coupling in armchair graphene nanoribbon junctions

*Suchun Li* [1,2,3], *Young-Woo Son* [4], *Su Ying Quek* [1,2, a)]

[1] Department of Physics, Faculty of Science, National University of Singapore, 2 Science Drive 3, Singapore 117551

[2] Institute of High Performance Computing, Agency for Science, Technology and Research, 1 Fusionopolis Way, #16-16 Connexis, Singapore 138632

[3] NUS Graduate School for Integrative Sciences and Engineering, National University of Singapore, 28 Medical Drive, Singapore 117456

[4] Korea Institute for Advanced Study, Seoul 130-722, Korea

______________________

[a)] Author to whom correspondence should be addressed.   Tel: (65) 6601 3640.   Fax: (65) 6777 6126

Email: phyqsy@nus.edu.sg

**Abstract**

**In recent years, bottom-up synthesis procedures have achieved significant advancements in atomically-controlled growth of several-nanometer-long graphene nanoribbons with armchair-shaped edges (AGNRs). This greatly encourages us to explore the potential of such well-defined AGNRs in electronics and spintronics. Here, we propose an AGNR based spin valve architecture that induces a large magnetoresistance up to 900%. We find that, when an AGNR is connected perpendicularly to zigzag-shaped edges, the AGNR allows for long-range extension of the otherwise localized edge state. The huge magnetoresistance is a direct consequence of the coupling of two such extended states from both ends of the AGNR, which forms a perfect transmission channel. By tuning the coupling between these two spin-polarized states with a magnetic field, the channel can be destroyed, leading to an abrupt drop in electron transmission.**

The prospect of all-carbon nanoelectronics has motivated significant interest in the transport of electrons through graphene and graphene nanoribbon (GNR) based junctions [1-3]. The weak intrinsic spin-orbit coupling in graphene also makes it an attractive candidate for replacing conventional materials in spintronics applications. Several interesting spin transport properties not found in other materials have been predicted. Most of these predictions, such as giant magnetoresistance [4,5] and half-metallic spin transport [6] have been centered on GNRs with zigzag atomic edges, with three-fold coordinated edge carbons. Such zigzag edges have spin-polarized edge states close to the Fermi energy [7,8]. On the other hand, significant progress has been made in the controlled atomic-scale synthesis of several-nanometer-long GNRs with *armchair* edges (AGNRs), all with specific widths [9-13]. Yet, to date, little is known about the potential of such well-defined AGNRs in electronics or spintronics. One notable prediction, later confirmed by experiments [14], showed that the application of an external magnetic field could reduce the quantum confinement effect in AGNRs, resulting in modified electrical resistance [15].

In this work, we predict, using first-principles calculations, that AGNRs can exhibit a large magnetoresistance, if connected to wider AGNR electrodes via transverse zigzag-edges on both ends. The huge magnetoresistance arises directly from channel formation due to coupling between spin-polarized interface-states with same energies on both sides of the junction. We show that this channel formation is an intrinsic property of the AGNRs − in particular, we find that for AGNRs of specific widths, this interface coupling is remarkably long-ranged. We further show that the essential physics of this channel remains the same in the presence of graphene or boron nitride substrates.

First principles transport calculations are performed using a two-lead system within the framework of density functional theory (DFT), using both SCARLET [16] and TRANSIESTA [17] for which consistent results are obtained. We use the local density approximation (LDA) and local spin density approximation (LSDA) for the exchange-correlation functional, as implemented in SIESTA [18], and a double-zeta basis-set, with a k-point sampling of $1\times1\times16$ for the lead calculation. All structures are fully relaxed until the forces on atoms are less than 0.01 eV/Å. For finite bias calculations, the wider AGNR was doped with 0.05% boron using the Virtual Crystal Approximation [19].

We refer to an AGNR with $n$ carbon atoms spanning its width (see Fig. 1(a)) as an $n$-AGNR. According to $n = 3p$, $3p+1$, or $3p+2$ (where $p$ is an integer), AGNRs can be classified into three families showing different electronic properties [20,21]. Without loss of generality, we begin our discussion with the charge transport properties of a prototypical device made of AGNRs in the $3p+2$ family − a middle 5-AGNR resistive part, connected on both sides to wider-width 23-AGNR electrodes, via transverse zigzag-edges (Fig. 1(a)). All the edge carbon atoms are passivated with a hydrogen atom. To prevent steric hindrance from hydrogen atoms at the interface, the middle-AGNR segment consists of an integer number of unit cells. Since this junction has zigzag-edge interfaces on both sides, we shall refer to it as the Z-Z structure. Electrons and holes that are non-spin-polarized will be transported through this Z-Z structure according to the black transmission curve in Fig. 1(c), with two resonant transmission peaks close to the Fermi level $E_F$, at -0.16 eV and 0.12 eV, respectively. The corresponding eigenchannel wavefunctions representing the nature of the conducting states at these two resonant peaks are mainly distributed at the two zigzag-edge interfaces and inside the middle 5-AGNR segment, suggesting that these states are related to the zigzag-edge interfaces.

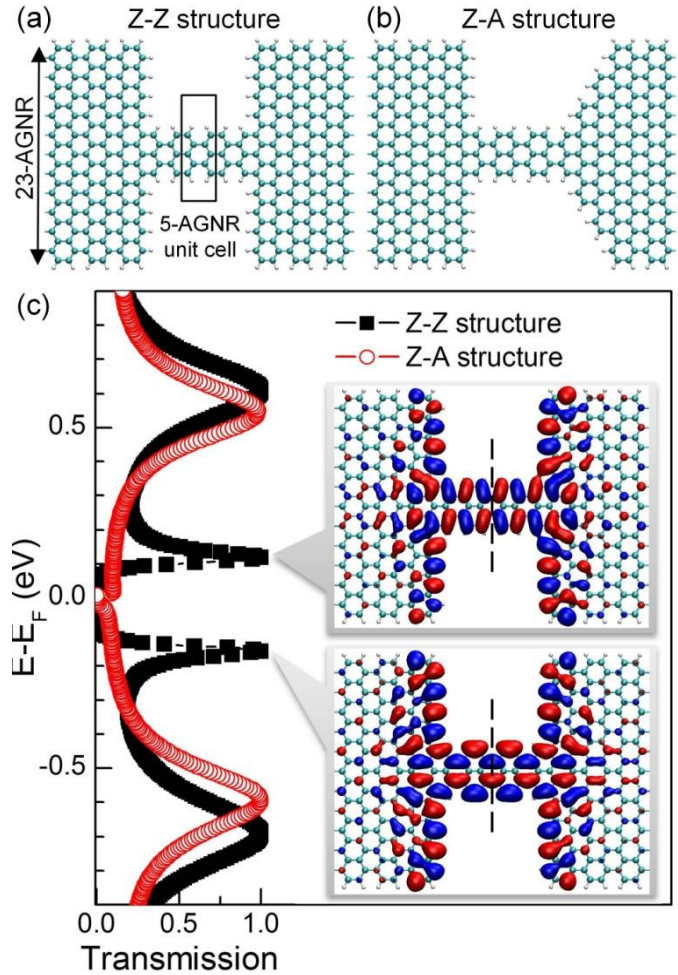

**FIG. 1** (a-b) atomic structures of an AGNR junction with two zigzag-edge interfaces (Z-Z structure) and one zigzag-edge interface plus one armchair-edge interface (Z-A structure). The length of the middle 5-AGNR segment is 3 unit cells for both structures. (c) transmission curves of junctions shown in (a) and (b). Inset of (c): real parts of the eigenchannel wavefunctions (isosurfaces with isovalue = +/- 0.025) at the two transmission peaks (the imaginary parts show the same features). The black dashed line indicate the plane normal to and through the center of the two-dimensional structure.

Interestingly, when one of the two zigzag-edge interfaces is replaced by armchair-edge interface (Z-A structure; Fig. 1(b)), the two resonant transmission peaks close to $E_F$ disappear (red curve in Fig. 1(c)). This observation clearly indicates that zigzag edges on both interfaces of the junction are necessary for these two peaks. In addition, we note that the occupied eigenchannel wavefunction (at -0.16 eV) is symmetric with respect to the plane normal to and through the center of the two-dimensional structure, as indicated by the black dashed line in Fig. 1(c) inset. On the other hand, the unoccupied eigenchannel wavefunction (at 0.12 eV) is antisymmetric with respect to the same plane. This observation of symmetry and antisymmetry in the two wavefunctions, together with the fact that both zigzag edges are required for the transmission peaks, strongly suggest that the states responsible for these peaks arise from bonding and antibonding combinations of two "original" states close to or at $E_F$ and related to the zigzag edges.

What could these "original" states be? To proceed, we use periodic boundary conditions to probe the electronic states close to and at $E_F$, which may not be evident from the transport calculations due to the non-conducting nature of these states or the gap of the AGNR leads. For the Z-Z structure, we observe two typical zigzag edge states [7,8] at $E_F$ (Supplementary Fig. S1(a)). These edge states are localized mainly at the zigzag edges, and decay towards the 23-AGNR, but *without* any extension into the region of the 5-AGNR segment. For the Z-A periodic structure, however, besides the usual zigzag edge state (Supplementary Fig. S1(b)), we observe an additional type of state at $E_F$, which is mainly distributed both at the zigzag edge and inside the entire 5-AGNR region, as shown in Fig. 2(a). Following its distribution property, we refer to this new state as a "zigzag + AGNR" state. Visually, it appears that the usual zigzag edge state extends seamlessly into the 5-AGNR segment without any spatial decay over the 5-AGNR region. States with similar pattern are observed when the zigzag edges at one interfaces of the Z-Z junction is replaced by other atomic structures, such as an $sp^3$-terminated zigzag-edge [22] (Fig. 2(b)). These results indicate that the "zigzag + AGNR" state at $E_F$ is likely to be robust against details of the geometry, as long as a single transverse sp$^2$ zigzag edge is interfaced with the 5-AGNR [23].

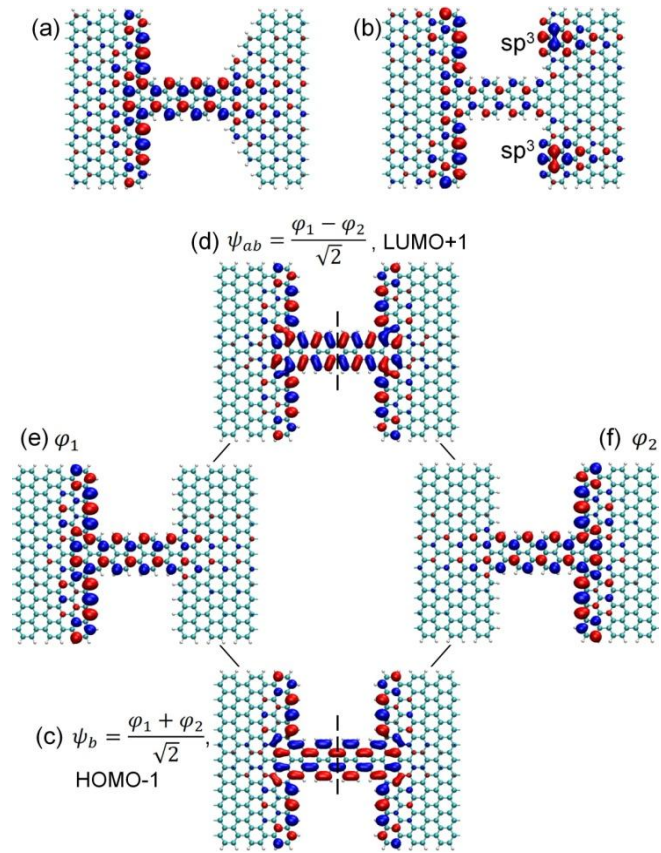

**FIG. 2** Wavefunctions of periodic AGNR systems (isosurfaces have isovalue = +/- 0.025). (a-b) Eigenwavefunctions at the Fermi energy. (c-d) The HOMO-1 and LUMO+1 eigenwavefunctions of the Z-Z structure, at energies of $E_F$ - 0.150eV and $E_F$ + 0.135eV, respectively. We refer to these states as bonding state $\psi_b$ and antibonding state $\psi_{ab}$. (e-f) Original "zigzag + AGNR" states deduced from the bonding and antibonding states by $\varphi_1 = (\psi_b + \psi_{ab})/\sqrt{2}$ and $\varphi_2 = (\psi_b - \psi_{ab})/\sqrt{2}$. Please note that $\varphi_1$ and $\varphi_2$ are not real eigenstates of the periodic Z-Z structure.

Going further away from $E_F$, we observe two states for the Z-Z structure, namely the HOMO-1 and LUMO+1 as shown in Fig. 2(c-d). These two states match very well in both energy and patterns to the transmission eigenchannel wavefunctions in Fig. 1(c), indicating that they are the states giving rise to the two transmission peaks near $E_F$. We had suggested earlier based on the symmetry and requirement on both zigzag edges that these states arise from bonding and antibonding combinations of two "original" states related to the zigzag edges. Here, we take the liberty to refer to the states in Fig. 2(c) and 2(d) as a bonding state $\psi_b$ and an antibonding state $\psi_{ab}$ respectively, and represent the two "original" states as $\varphi_1$ and $\varphi_2$, i.e.

$$\psi_b = \frac{\varphi_1 + \varphi_2}{\sqrt{2}} \qquad (1)$$

$$\psi_{ab} = \frac{\varphi_1 - \varphi_2}{\sqrt{2}} \qquad (2)$$

Under this hypothesis, we use equations (1) and (2) to deduce the two "original" states $\varphi_1$ (Fig. 2(e)) and $\varphi_2$ (Fig. 2(f)). Interestingly, $\varphi_1$ is the "zigzag + AGNR" state observed at $E_F$ in Fig. 2(a-b), and $\varphi_2$ is just the mirror-reflection of $\varphi_1$ corresponding to the other zigzag-edge interface. This confirms our hypothesis that HOMO-1 and LUMO+1 are bonding and antibonding couplings of two original "zigzag + AGNR" states. It is interesting to note that $\varphi_1$ and $\varphi_2$ are localized on different carbon sub-lattices in the 5-AGNR segment, since the 5-AGNR segment contains an integer number of unit cells. Coupling between $\varphi_1$ and $\varphi_2$ therefore couples the two sub-lattices.

Since the zigzag edge state is spin-polarized [7,8], the "zigzag + AGNR" state is also spin-polarized, with magnetic moments localized mainly on the zigzag edges. In zigzag graphene nanoribbons, the spin up and spin down edge states are split in energy by ~0.5eV according to first principles calculations [21], and larger when many-electron effects are taken into account [24]. Likewise, we expect here that each "zigzag + AGNR" state $\varphi_1$ ($\varphi_2$) is split into two states with opposite spins $\varphi_1^\uparrow$ ($\varphi_2^\uparrow$) and $\varphi_1^\downarrow$ ($\varphi_2^\downarrow$), where the one with majority spin is shifted down in energy by the magnetic exchange energy term $E_M$ and the other with minority spin is shifted up by the same amount $E_M$. Crucially, because good coupling between these states requires them to be at the same energy, we expect that the spin orientation at the two zigzag edge interfaces can be used to control the coupling between the states, thereby closing or opening the channels for electron transmission. When the spin orientations at both interfaces are parallel (P configuration), the spin up (down) original states on both sides of the junction will still be at the same energy, and therefore can couple equally well as in the non-spin-polarized case. In contrast, when the spin at one zigzag edge interface is pointed in the opposite direction as the spin at the other zigzag edge interface (antiparallel (AP) configuration), the spin up (down) original states at both sides of the junction will be at different energies, resulting in significantly reduced coupling and electron transmission.

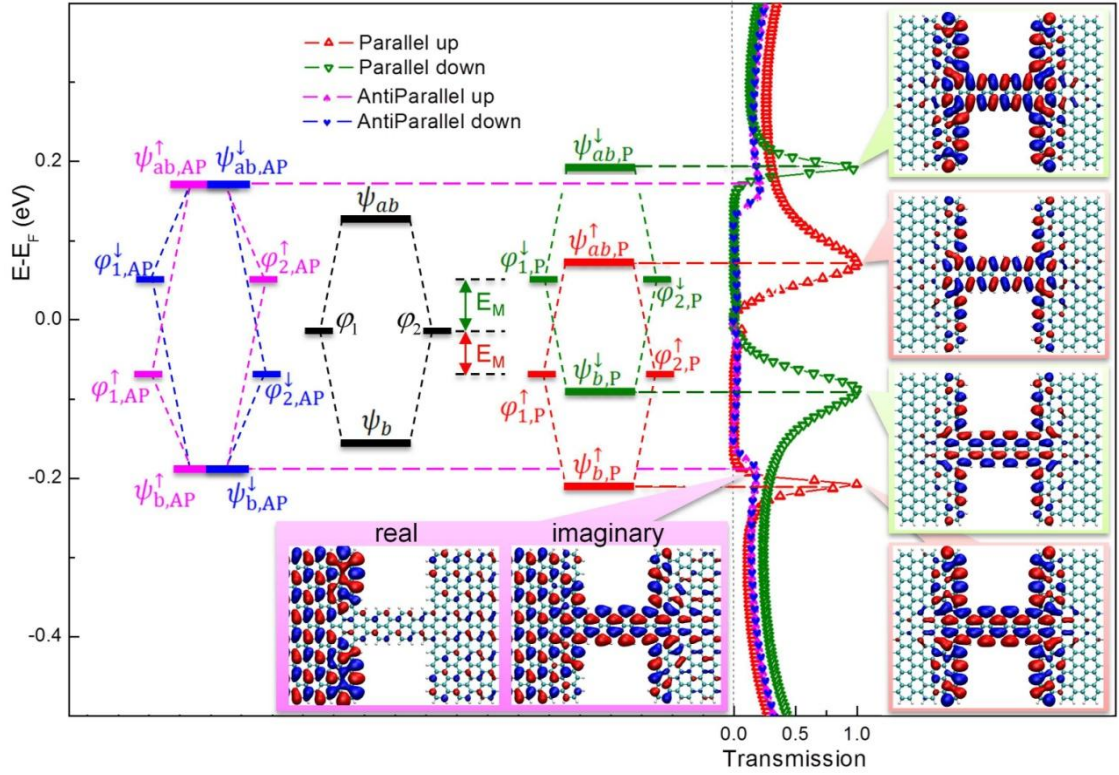

**FIG. 3** Spin-polarized transmission curves of the Z-Z junction with Parallel (P) and Antiparallel (AP) spins on zigzag-edges of two interfaces. Inset schematic: energy level diagrams illustrating the bonding and antibonding couplings of two original states in non-spin-polarized (middle black diagram), P (right red/green diagram), and AP (left pink/blue diagram) cases. Inset wavefunction isosurfaces (isovalue = +/- 0.025) on right side of transmission curve: real parts of eigenchannel wavefunctions at the four perfect transmission peaks for P case. Inset wavefunction isosurfaces (isovalue = +/- 0.005) on left side of the transmission curve: spin up eigenchannel wavefunction (real and imaginary parts) incident from the left at the bonding (loosely defined) peak of AP case.

The above hypothesis, illustrated in the energy level diagrams in Fig. 3 inset, is clearly supported by our first principles spin-polarized transmission results (Fig. 3). For the P configuration, the computed spin up (majority) and spin down (minority) transmission both show two resonant peaks with transmission ~1 close to $E_F$ (Fig. 3, red/green curves), but shifted down and up in energy relative to the non-spin-polarized case, respectively. The eigenchannel wavefunctions at these perfect transmission peaks (Fig. 3 inset, right side) are exactly the bonding and antibonding states. In particular, when we construct the "original" states from parallel spin-polarized eigenstates of the periodic structure by $\varphi_{1,P}^{\uparrow} = \frac{\psi_{b,P}^{\uparrow}+\psi_{ab,P}^{\uparrow}}{\sqrt{2}}$ ( $\varphi_{1,P}^{\downarrow} = \frac{\psi_{b,P}^{\downarrow}+\psi_{ab,P}^{\downarrow}}{\sqrt{2}}$ ) and $\varphi_{2,P}^{\uparrow} = \frac{\psi_{b,P}^{\uparrow}-\psi_{ab,P}^{\uparrow}}{\sqrt{2}}$ ( $\varphi_{2,P}^{\downarrow} = \frac{\psi_{b,P}^{\downarrow}-\psi_{ab,P}^{\downarrow}}{\sqrt{2}}$ ), we obtain essentially the same "original" states as in the non-spin-polarized case (Supplementary Fig. S2). On the other hand, the corresponding transmission peaks are significantly suppressed in the AP configuration (Fig. 3, pink/blue curves). The corresponding eigenchannel wavefunction (Fig. 3 inset, left side) indicates that electrons coming from the left lead are reflected at the first (real part) or second (imaginary part) interfaces they encounter.

The large difference in transmission spectra for P and AP configurations suggests that a large magnetoresistance (MR) can be achieved in spin valve architectures based on this structure. Indeed, our self-consistent finite bias calculations indicate that the MR (defined as $\frac{I_P - I_{AP}}{I_{AP}} \times 100\%$) can reach values as large as ~900% with a voltage bias of 0.3V (Fig. 4(b)). The fact that MR peaks close to 0.3V is consistent with the simple picture that the current in the P configuration would increase significantly as the bias window is enlarged, while the current in the AP configuration would remain very small until the small shoulders at ±0.2eV in the transmission spectra (Fig. 3 pink/blue curves) enter the bias window at bias ~ 0.4V (Fig. 4(a)).

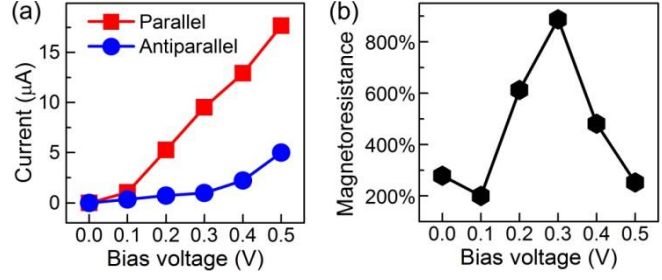

**FIG. 4** (a) IV curves of the Z-Z junction with P and AP configurations. (b) Magnetoresistance of the junction at various bias voltages.

Finally, we note that the coupled states $\psi_b$ and $\psi_{ab}$ are quite robust against various modifications to the structures. Firstly, they persist in the presence of graphene and boron nitride substrates (Supplementary Fig. S3). This can be understood since the two interesting states are originated from the π-electrons, which do not couple strongly to these substrates. Secondly, they persist when the middle AGNR is transversely shifted across the width. Thirdly and most interestingly, they persist in structures with *remarkably long* (can be infinite) middle AGNRs at almost the same energies due to the non-decaying feature [25] (See Supplementary Fig. S4). We have also observed such long-range coupling mediated by AGNRs of other widths, such as 11-AGNR and 17-AGNR (Supplementary Fig. S5), both with width $n = 3p+2$. For AGNRs in other families, we also observed the formation of bonding and antibonding states close to the Fermi level, but the coupling is not always long-ranged. Since AGNRs in different families have different energy gaps and wavefunctions of different patterns, this family-dependent behaviour is consistent with our understanding that it is the intrinsic property of the narrower AGNR region that allows a long-range spatial extension of the zigzag edge state.

In summary, we have predicted very large magnetoresistance in nanostructured armchair graphene nanoribbons in spin valve architectures. This large MR arises from the intrinsic ability of the middle AGNR region to couple spin-polarized interface states with the same energies on both sides of the junction. In the wake of recent experimental advancements in the bottom-up synthesis of atomically well-defined AGNRs [12,13] as well as progress toward nanostructuring in graphene [26-30], our predictions are timely and pave the way for further exciting discoveries and potential spintronic applications of AGNR-based structures.

S.Y.Q. gratefully acknowledges Y.P. Feng for mentorship and freedom to pursue this project with S.L. S.Y.Q. and S.L. thank Y.P. Feng and C.K. Gan for helpful discussions, and A*STAR for funding via the A*GS Scholarship and IHPC Independent Investigatorship respectively. S.Y.Q. also thanks the Singapore NRF for funding via the NRF Fellowship (NRF-NRFF2013-07). Y.-W. S. was in part supported by the NRF grant funded by Korea MEST (QMMRC, R11-2008-053-01002-0 and Nano R&D program 2008-03670). We thank A*CRC and NUS Graphene Research Centre for computational support and H.J. Choi for the SCARLET code.